\documentclass[preprint,longnamesfirst]{aastex}

\newcommand{\thisdir}{.}
\newcommand{\kpc}{\mathop{\rm kpc\,}\nolimits}
\newcommand{\Mpc}{\mathop{\rm Mpc\,}\nolimits}
\newcommand{\Msun}{\mathop{\rm M_{\odot}\,}\nolimits}

\newcommand{\fig}{Fig.~\ref}

\newcommand{\tab}{Table~\ref}

\newcommand{\sect}{Sec.~\ref}

\newcommand{\Hydra}{{\sc hydra}}

\newcommand{\expd}[1]{\times 10^{#1}}
\newcommand{\mean}[1]{\langle #1 \rangle}

\newcommand{\fraction}[2]{\mbox{\scriptsize$^{{#1}\!}/_{\!{#2}}$}}

\ifx\undefined\onequarter
 \newcommand{\onequarter}{\fraction{1}{4}}
\else
 \renewcommand{\onequarter}{\fraction{1}{4}}
\fi
\ifx\undefined\onethird
 \newcommand{\onethird}{\fraction{1}{3}}
\else
 \renewcommand{\onethird}{\fraction{1}{3}}
\fi
\ifx\undefined\onehalf
 \newcommand{\onehalf}{\fraction{1}{2}}
\else
 \renewcommand{\onehalf}{\fraction{1}{2}}
\fi
\ifx\undefined\twothirds
 \newcommand{\twothirds}{\fraction{2}{3}}
\else
 \renewcommand{\twothirds}{\fraction{2}{3}}
\fi
\ifx\undefined\threequarters
 \newcommand{\threequarters}{\fraction{3}{4}}
\else
 \renewcommand{\threequarters}{\fraction{3}{4}}
\fi

\ifx\undefined\fourthirds
 \newcommand{\fourthirds}{\fraction{4}{3}}
\else
 \renewcommand{\fourthirds}{\fraction{4}{3}}
\fi
\ifx\undefined\threehalfs
 \newcommand{\threehalfs}{\fraction{3}{2}}
\else
 \renewcommand{\threehalfs}{\fraction{3}{2}}
\fi
\ifx\undefined\threehalves
 \newcommand{\threehalves}{\threehalfs}
\else
 \renewcommand{\threehalves}{\threehalfs}
\fi

\newcommand{\figscaleone}{1}

\begin{document}

\title{Hierarchical clustering and the baryon distribution in galaxy clusters}

\author{Eric R. Tittley and H. M. P. Couchman\footnote{Present address: McMaster University, Department of Physics and Astronomy, Hamilton, Ontario, L8S 4M1, Canada}}
\affil{University of Western Ontario}
\affil{
Department of Physics and Astronomy
London, Ontario, N6A 3K7, Canada
}
\email{etittley@astro.uwo.ca}
\email{couchman@physics.mcmaster.ca}

The baryon fraction of galaxy clusters in numerical simulations is found to be dependant on the cluster formation method.  In all cases, the gas is anti-biased compared with the dark matter.  However, clusters formed hierarchically are found to be more depleted in baryons than clusters formed non-hierarchically.  There is a depletion of 10 to 15\% for hierarchically formed clusters while the depletion is less than 10\% for those clusters formed non-hierarchically.  This difference is dependent on the mass of the clusters.
The mean baryon enrichment profile for the hierarchically formed clusters shows an appreciable baryon enhancement around the virial radius not seen in the clusters formed without substructure. If this phenomenon applies to real clusters, it implies that determinations of the baryon fractions in clusters of galaxies require data extending beyond the virial radius of the clusters in order to achieve an unbiased value.

\section{INTRODUCTION}
\label{sec.Intro}
Since the early hydrodynamic simulations of galaxy clusters \citep{Evrard90}, the global distribution of the baryons has been of interest.  In particular, the local enhancement or deficit of the baryons compared with the dark matter has profound implications for inferences of the universal baryon fraction as parametrized by $\Omega_b$.  Recall that $\Omega$ is the ratio of the mass-energy density of the universe to the critical density required to close the universe.  The ratio, $\Omega_b$, is the contribution of the baryonic component to $\Omega$, while $\Omega_m$ is the contribution of all mass.  The fraction, $\Omega_b$, is constrained by primordial nucleosynthesis calculations which are sensitive to the cosmological model.  Currently, estimates vary between a reported low of $\Omega_b=0.013\pm 0.003 h^{-2}$ \citep{WF95} to $\Omega_b=(0.020\pm 0.002) h^{-2}$ \citep{Bludman98}.  Since galaxy clusters are the largest objects in the universe for which one can observe the baryon content as well as derive the total mass, they provide the most unbiased samples from which to calculate the baryon fraction of the universe.  The values for $\Omega_b$ and the baryon fraction, $f_b\equiv \fraction{\Omega_b}{\Omega_m}$, derived from observations of clusters, may be combined to derive $\Omega_m$, the total contribution of matter to $\Omega$.  Observations of this sort find baryon fractions of 10 to 22\% \citep{WF95,WJF97}.  This implies $\Omega_m<0.9$ for $f_b=0.1$ and $h=0.5$ while suggesting it is probably closer to 0.3 if we take $f_b=0.15$ and $h=0.65$.  However, the derivation of $f_b$ from the observations of clusters assumes the dark matter and hot gas are distributed in constant proportions within the cluster.  Consequently, an understanding of the concentration factor of the gas in clusters is required since it represents the degree of biasing. As well, this biasing can have consequences in regards to constraining the deceleration parameter, $q_0$ \citep{RFJ98}.

Numerical simulations generally agree that the gas is anti-biased in clusters of galaxies \citep{Evrard90,TC92,CO93a,Kang94,ME94,PTC94,NFW95,AN96,Lubin96,PEG96}. Results have been produced to the contrary \citep{OV97} using two-dimensional codes. \citet{PTC94} explains the anti-bias as a result of the merging process in which gas is shocked, permanently removing energy from the dark matter component and passing it to the gas.

However, there is an uncertainty in the actual amount of bias.  The bias between the universal baryon fraction and that found in clusters cannot be measured directly but must be inferred from numerical simulations.  The numerical simulations must assume a model for the universe.  Since the actual cosmogony of the universe is unknown in the details, variation among baryon fractions found in different model universes adds uncertainty.  A better understanding of how this bias depends on cosmological models would assist in understanding the actual baryon fraction of the universe. 

In hierarchical clustering, the largest structures forming at a given time do so via the amalgamation of many smaller structures which have formed at an earlier time.  This is owing to the form of the initial density perturbation spectrum in which small-scale perturbations have higher initial amplitudes than large-scale. In contrast, non-hierarchical clustering involves structure formation from the collapse of large structures with smooth density distributions.  The results of numerical simulations compared with observations support the theory that we live in a universe in which structure is formed hierarchically.  The degree to which the hierarchical nature affects the baryon distribution in galaxy clusters is not entirely clear.  In this paper, a comparison of the baryon fraction bias will be made using numerical simulations of galaxy clusters formed hierarchically and non-hierarchically. By using these extreme cases, the significance of hierarchical clustering itself will be determined.  The analysis will concentrate not on the properties of individual clusters, but on the global mean properties of scaled quantities.

The layout of the paper is as follows. In \sect{Sec.Sims}, the simulations are described. A brief description of the analysis methods are given in \sect{Sec.Analysis}. The effect of hierarchical clustering on the baryon fraction is examined in \sect{sec.BF}.  The results are discussed in \sect{sec.Conclusions}.

\section{SIMULATIONS}
\label{Sec.Sims}
The simulations used for this study are the same as described in \citet{ET99Hier}. They comprise a set of five high resolution simulations in which hierarchical clustering was reduced or eliminated in four of the simulations via smoothing of the density perturbations in the initial conditions file.

The initial, unsmoothed density distribution was scale-invariant, with a spectral index of $n=-1$, such that the density fluctuation power-spectrum had the form of the power law, $P(k)\propto k^n$.  It was normalised to set the variance on the scale of $8h^{-1}\Mpc$ to $\sigma_8=0.935$.

The smoothing was done using a top-hat method for two of the simulations and a low-bandpass filter for the other two.  For the top-hat method, smoothing over radii of $3h^{-1}\Mpc$ and $7h^{-1}\Mpc$ was done.  The frequency cutoffs were $2\pi(7h^{-1}\Mpc)$ and $2\pi(14h^{-1}\Mpc)$.  Subsequent references to these simulations will omit the $h^{-1}$ and $2\pi$ factors. The Hubble constant parameter, $h=0.65$, was adopted for these simulations. The details of the simulations are given in \tab{tab.Sim.simulations}.

The simulations contained $2\times 64^3$ particles, half of which were collisionless particles representative of the dark matter component while the other half were collisional which represented a gas phase.  The particles had masses of $9.27\expd{10}\Msun$ for the dark matter and $1.03\expd{10}\Msun$ for the gas. The total mass density is sufficient to produce a flat cosmology, without a cosmological constant.

The matter was evolved in a box with periodic boundary conditions on a scale of $40 h^{-1} \Mpc$ in co-moving coordinates.  The N-body ${\rm AP}^3{\rm M}-{\rm SPH}$ code, \Hydra\ \citep{CTP} was used for all simulations.  Two-body interactions were reduced by the use of gravitational softening.  This was provided by the softening parameter, $\epsilon=20 h^{-1}\kpc$. Cooling was neglected.  The cooling time for the bulk of the cluster gas is estimated to be well over the age of the clusters.

\section{ANALYSIS}
\label{Sec.Analysis}
The same sample of clusters as selected in \citet{ET99Hier} was used here. The details of the sets of clusters for each simulation are given in \tab{tab.analysis.clusters}. Clusters were selected by a tomographic deprojection method.

For each cluster, masses and radii were calculated at overdensities of 200 and 500. The radius for an overdensity of 200, $R_{200}$, is the radius of a cluster interior to which the mean density is 200 times the background density, which for these simulations is the critical density, $\rho_c$. A similar definition holds for $R_{500}$.  The mass interior to these radii are denoted, $M_{200}$ and $M_{500}$, respectively.  The overdensity of 200 was chosen since it is on the conservative side of 178, the overdensity at which a spherically symmetric sphere of matter virialises, according to analytic models of top-hat collapses.
The masses of both the dark matter, $M_{dark}$, and gas, $M_{gas}$, were found at this virial radius, along with the total mass, $M_{200}$.

A lower limit to the size of the clusters was set by the requirement that each cluster, within the overdensity radius of $R_{500}$, have at least 300 gas particles and 300 dark matter particles.  This ensures the densities are calculated correctly \citep{ET99Hier}.

Profiles of the dark matter and gas densities were calculated for each of the clusters.  This was done by summing the mass contributions of the particles falling in exponentially separated radial bins centred on the clusters. The profiles were scaled by the approximate virial radius, $R_{200}$.

Mean density profiles for each simulations were then calculated from the profiles of the individual clusters in each simulation.  For the calculation of the mean density profile, the individual cluster profiles were weighted by their respective mass, $M_{200}$.  This weighting removes the bias towards the more plentiful low-mass clusters.  However, for the density profiles, this weighting was found to have only a small effect, demonstrating the lack of dependency on mass for the profiles.

\section{BARYON FRACTION}
\label{sec.BF}
\subsection{The concentration parameter, $\Upsilon$}
\label{sec.BF.Upsilon}
If we define $\Upsilon_{200}$ as the gas fraction in a spherical region
of overdensity 200 normalised to the cosmic value, then we have
\citep{WNEF}
\begin{equation}
\Upsilon_{200} \equiv \frac{M_{gas}}{M_{gas}+M_{dark}}\frac{\Omega_m}{\Omega_b}.
\label{eq.BF.Upsilon}
\end{equation}

Similarly, we may define $\Upsilon(r)$ as the normalised gas fraction in a shell
of radius $r$.  Here, $r$ is replaced by the dimensionless parameter $r/R_{\bar{\delta}}$ since radial profiles of both the dark matter and gas densities  are characterised by the virial radius.

\subsection{Variation of $\Upsilon$ on cluster mass}
\label{sec.BF.vsMass}
\begin{figure}
\epsscale{\figscaleone}
\plotone{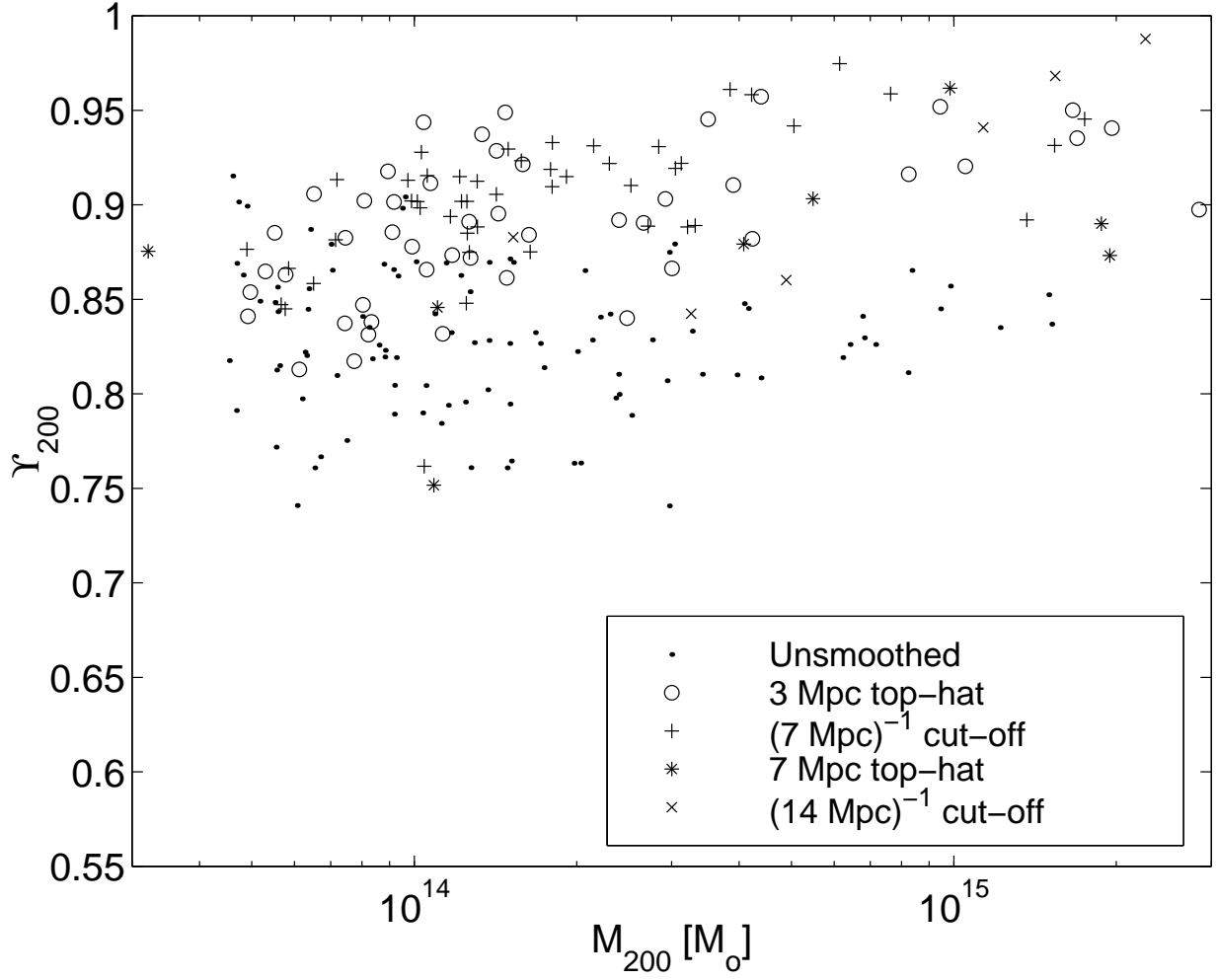}
\caption[The normalised baryon fraction as a function of mass]{
The normalised baryon fraction as a function of mass. The baryon fraction is measured by $\Upsilon_{200}$ as a function of $M_{200}$.}
\label{fig.Baryon.BF.Upsilon}
\end{figure}

The values of $\Upsilon_{200}$, when plotted versus $M_{200}$ (\fig{fig.Baryon.BF.Upsilon}), support the values found in \citet{Evrard97}~and~\citet{ENF98} of $\Upsilon_{200}=0.85$ to $0.90$ if the sample is restricted to the largest clusters.

For the lower mass clusters, the clusters in the unsmoothed model have values ranging down to $0.75$.  There is no trend with mass for this model other than an increase in dispersion.  The mean for the hierarchically formed clusters is $0.84$ with the standard deviation varying from $\pm 0.04$ for clusters less massive than $10^{14} \Msun$ to $\pm 0.01$ for clusters more massive than $10^{15} \Msun$.

For the smoothed models, the trend is for $\Upsilon_{200}$ to decrease with mass with little change in the dispersion.  For clusters more massive than $10^{15} \Msun$, $\Upsilon = 0.87$ to $0.97$.  Below $10^{14} \Msun$, this drops down to the range $0.82$ to $0.92$.

\subsection{Baryon concentration profiles}
\label{sec.BF.profiles}
Since it is found that both the dark matter and gas density profiles scale with the virial radius, it stands to reason that the radial profile of the baryon concentration should also scale with radius.  Numerical simulations of galaxy clusters using a 1-D Lagrangian code indicate that this is not true.  \citet{KP97} reports that $f_b$ varies by a factor of about two at a given scaled radius between clusters that differ in mass by two orders of magnitude, with the less massive cluster having the smaller baryon fraction.  This trend is consistent with the variation in $\Upsilon_{200}$ with the mass of the cluster seen for the clusters formed from smoothed initial conditions.  However, since this factor is comparable to the cluster to cluster variation, it is justified once again to look at the mean radial profile of all the clusters in a model.

\begin{figure}
\epsscale{0.7}
\plotone{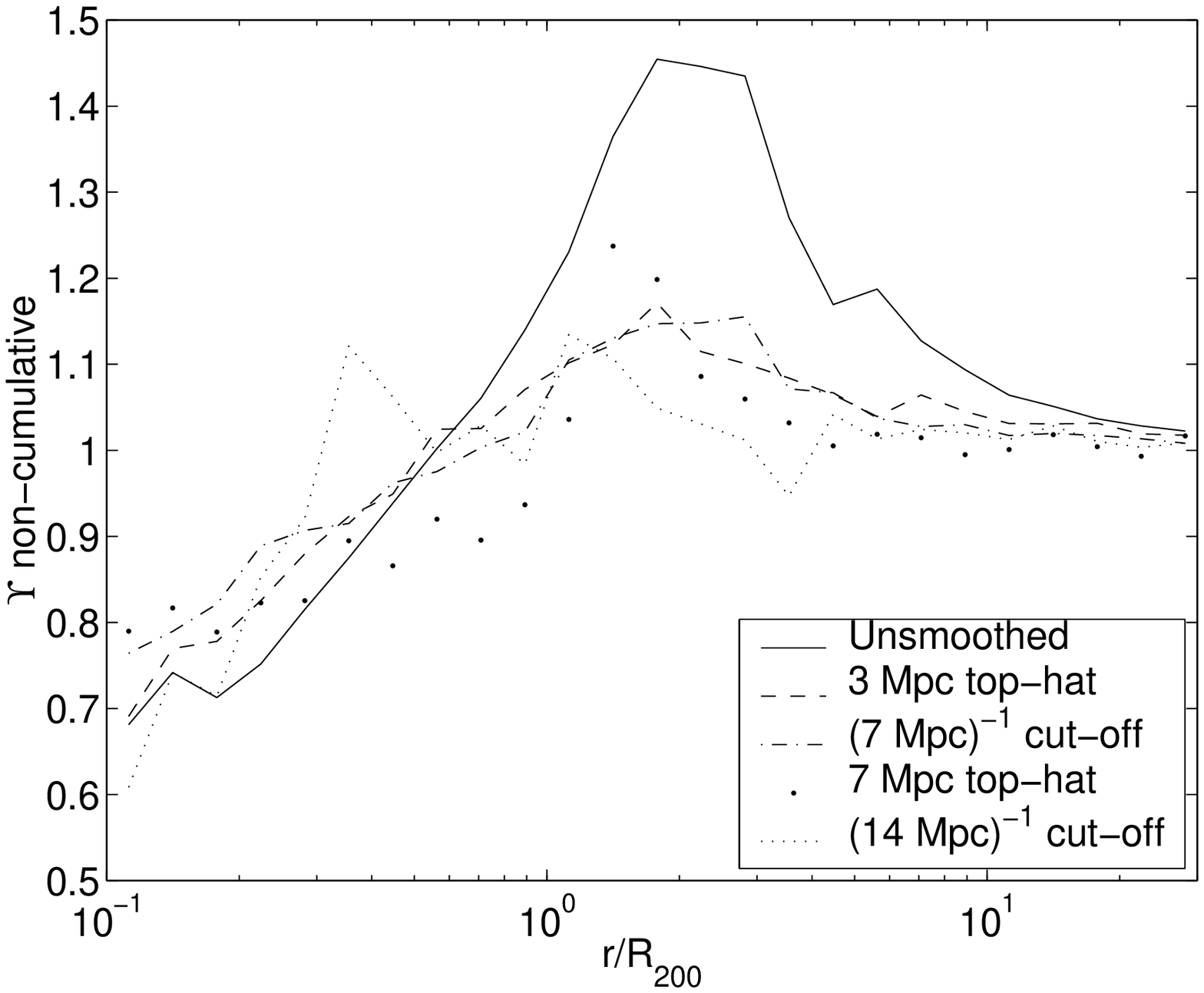}
\plotone{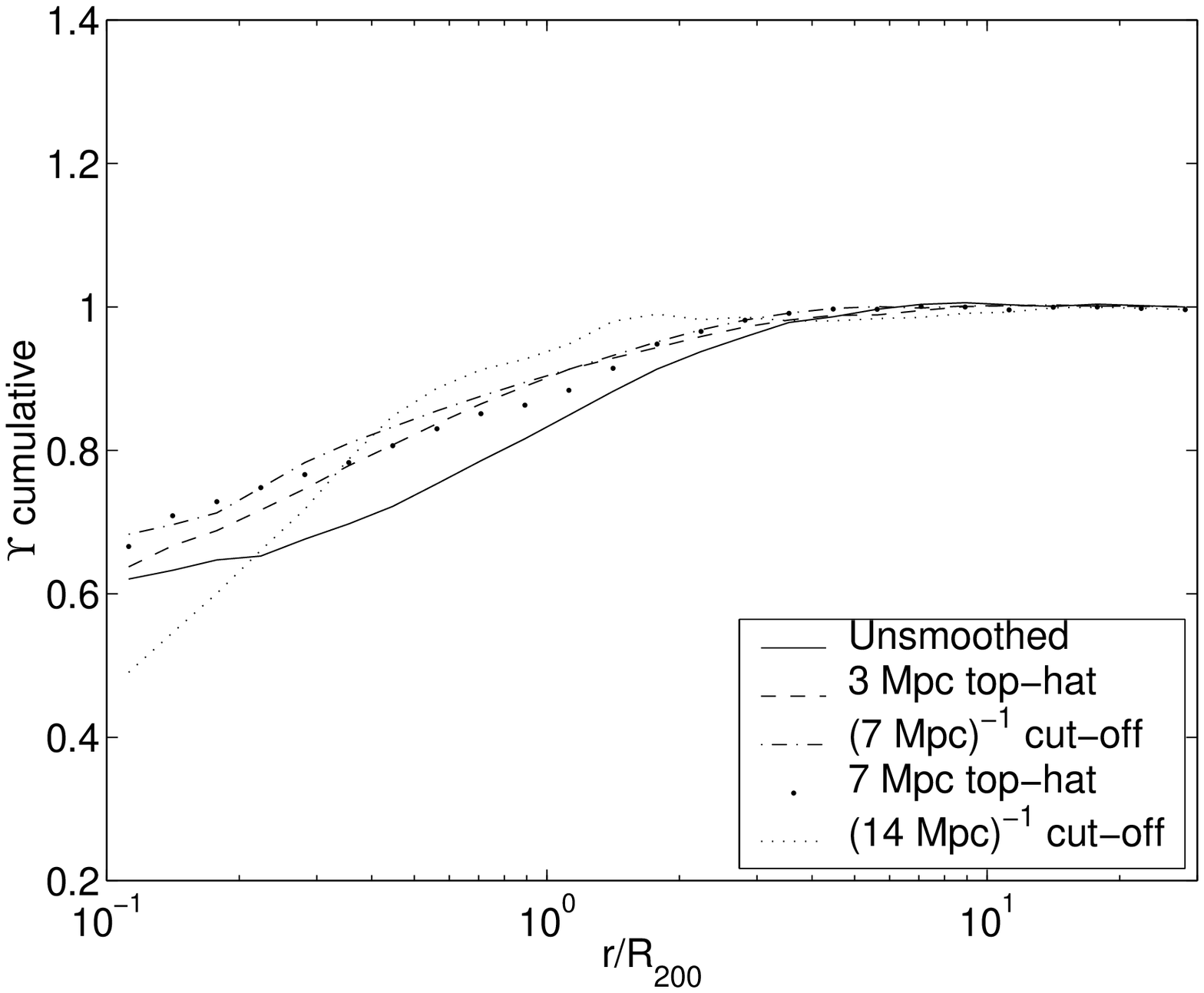}
\caption{
The normalised mean baryon fraction profiles, as measured by $\Upsilon(r/R_{200})$,  for the models. At the top are the profiles for $\mean{\Upsilon(r/R_{200})}$ at the specified radii.  The mean cumulative enrichment profiles are given in the lower panel.
}
\label{fig.Baryon.Prof.BF}
\end{figure}

The baryon fraction profiles vary considerably for those clusters formed in the
simulation with hierarchical clustering from those formed otherwise (\fig{fig.Baryon.Prof.BF}, top) particularly around the virial radius.  The baryon fraction is enhanced in a much deeper region ($0.7 < r/R_{200} < 10$) in the hierarchical case than in the case with smoothed initial conditions ($1 < r/R_{200} < 4$) and the enhancement is almost 4 times greater.  In both cases, the peak in the baryon enrichment occurs just beyond the virial radius. Towards the more central regions, the enrichment in the hierarchical clusters drops off slightly more steeply.

The discrepancies are not so large for the cumulative profile (\fig{fig.Baryon.Prof.BF}, bottom).  The clusters in the hierarchical model are generally more depleted in baryons than those in the smoothed models, in agreement with the results in \sect{sec.BF.vsMass}. The mean profile for the unsmoothed model compares well those of the simulated clusters discussed in \citet{ENF98}.  Those are simulated for a low-density universe.

It must be noted that there are large deviations in these profiles among the many clusters in the hierarchical scenario. These mean profiles represent trends. For the clusters formed non-hierarchically, however, the baryon profiles are fairly consistent among clusters.

Interior to $1.5 R_{200}$, the mean $\Upsilon$ profile for the unsmoothed model may be fit by $\Upsilon(r) = 0.67\pm 0.03 + (0.52\pm 0.05) \fraction{r}{R_{200}}$, non-cumulative.  For the mean of the profiles for the non-hierarchically formed clusters, the fit takes a shallower form, $\Upsilon(r) = 0.76\pm 0.02 + (0.30\pm 0.07)\fraction{r}{R_{200}}$.

\section{DISCUSSION}
\label{sec.Conclusions}
Using a series of simulations in which clusters were formed either hierarchically and non-hierarchically, the relevance of hierarchical structure formation to the baryon distribution in clusters was explored.  Non-hierarchical clustering was achieved by smoothing the initial density distribution.

It is found that the baryons are anti-biased in clusters of galaxies and this biasing is dependent on the presence of smoothing in the initial conditions.  The bias parameter $\Upsilon \simeq 0.85$ for the hierarchically formed clusters.  For the clusters formed from smoothed initial conditions, $\Upsilon \simeq 0.92$, indicating less of an anti-bias. For the non-hierarchically formed clusters, the normalised baryon fraction is dependent on the mass of the cluster, as well, which is not observed for the hierarchically formed clusters. 

In all cases, the baryon fraction was found to increase from the centre of the clusters, outward.  Interior to $0.1 R_{200}$, the normalised factor spans $\Upsilon=0.5$ to $0.7$. We determined that the baryon fraction profile, as parametrized by the normalised fraction, is steeper for hierarchically formed clusters.  For these, it was found that $f_b \propto (0.52\pm 0.05)r$ interior to $1.5 R_{200}$.  For non-hierarchically formed clusters, the profiles were shallower, regardless of the degree of initial smoothing, with a mean dependency of $f_b \propto (0.30\pm 0.07)r$. The baryon fractions peak between 1 to 2 virial radii.  For the clusters formed hierarchically, the baryons `pile up' in a deep region spanning $0.7 < r/R_{200} < 10$.  This implies that measurements of the baryon fraction of clusters are sensitive to the radius outward to which the baryon content is integrated.

Baryon fraction profiles determined from observations of clusters via the deprojection of the gas density \citep{WJF97} support the result that the baryon fractions increase with increasing radius \citep{MV97}. However, the profiles are often shallower than those found here.  \citet{WF95} finds from a sample of 19 clusters that the cumulative measurement of $f_b$ increases with radius following roughly the $f_b=0.06 R$ with $R$ in $\Mpc$.  Similarly, \citet{WJF97} finds a fit of $f_b=(0.12\pm0.04)R^{0.7\pm0.3}$ for a sample of 28 clusters without cooling flows. With cooling flows, a higher central value for $f_b$ and a shallower profile are reported, as would be expected.  The fractional slopes of both these fits are steeper than that of the mean $\Upsilon$ profile of all the models. This is likely due to the lack of cooling in the simulations, which would deposit a concentrated amount of gas near the cores of the clusters.  This gas would represent a small fraction of the total mass of the halo gas, but would elevate the baryon fraction in the interior. This should not greatly affect the results at radii near the virial radius, since the cooled gas would constitute a small fraction of the total gas mass at that point.

These results indicate that the observed baryon fractions in galaxy clusters lead to under-estimates of the universal baryon fraction by a factor of $5\%$ to $20\%$ and that this factor is dependent on the cosmology.  The anti-biasing of the gas is greatest in the hierarchical run, consistent with the picture of \citet{PTC94} in which the bias is due to transfer of energy from the dark matter to the gas during merging.  Since cooling flows are not created in these simulations, the true bias factor is not well established.  However, cooling flows will only alter the baryon fractions in the inner radii since the total mass cooled is only on the order of $10^{12} \Msun$ or less.  Ultimately, this result exacerbates the baryon overdensity problem in galaxy clusters.  The variation in the enrichment of baryons with distance from the cluster centres also makes measurements of $f_b$ problematic.  It is crucial that the cluster be observed to radii of up to two virial radii to reduce the error to less than $10\%$.

Observations of the baryon fraction in clusters will be hampered by this biased distribution of gas in clusters.  Measurements must be made to more than twice the virial radius, sampling regions for which accurate densities of both mass and gas are difficult. This uneven distribution is not as significant if there is any degree of smoothing, implying that any attempt to formulate a correction factor will be dependent on the cosmology at the $10\%$ level.  Fortunately, this is at the present level of measurement error.

\bibliographystyle{plainnat_mnras}
\bibliography{biblio}

\renewcommand{\thisdir}{Tables}

\begin{table}
\begin{center}
\begin{tabular}{ccc}
\hline
Smoothing method        & $r_{smooth}$  & Renormalised   \\
                        &($h^{-1} \Mpc$)& to $\sigma_8$\\
\hline
unsmoothed 		& 0             &  N.A.\\
$3\Mpc$ top-hat		& 3             &  Yes\\
$(7\Mpc)^{-1}$ k-cutoff	& 7             &  Yes\\
$7\Mpc$ top-hat		& 7             &  No\\
$(14\Mpc)^{-1}$ k-cutoff& 14            &  No\\
\hline
\end{tabular}
\end{center}
\caption{
Properties of the simulations. Given are the effective smoothing radius and whether the data, after smoothing, were renormalised to $\sigma_8=0.935$.
}
\label{tab.Sim.simulations}
\end{table}

\begin{table}
\begin{center}
\begin{tabular}{lccc}
\hline
                        & $N_{clusters}$& $M_{min}[10^{15}\Msun]$  	& $M_{max} [10^{15}\Msun]$ \\
\hline
unsmoothed	        & 100		&  0.05				& 1.52 \\
$3\Mpc$ top-hat         & 48 		&  0.05			    	& 2.85 \\
$(7\Mpc)^{-1}$ k-cutoff & 48 		&  0.05				& 1.75 \\
$7\Mpc$ top-hat         & 8  		&  0.03			   	& 1.95 \\
$(14\Mpc)^{-1}$ k-cutoff& 6 		&  0.15				& 2.27 \\
\hline
\end{tabular}
\end{center}
\caption[Results of the cluster search]{
Results of the cluster search.  Given for each run is the number of clusters found, $N_{clusters}$, and the range of masses of the clusters, $M_{min}$ and $M_{max}$.}
\label{tab.analysis.clusters}
\end{table}

\renewcommand{\thisdir}{Figures}

\end{document}